# *InsteaDMatic:* Towards cross-platform automated continuous rotation electron diffraction

Authors

**Maria Roslova[a], Stef Smeets[a,1], Bin Wang[a], Thomas Thersleff[a], Hongyi Xu[a] and Xiaodong Zou[a]***

[a]Department of Materials and Environmental Chemistry (MMK), Stockholm University, Svante Arrhenius väg 16C, Stockholm, SE-10691, Sweden

Correspondence email: x.zou@mmk.su.se

[1]Present address: Department of Bionanoscience, TU Delft, Van der Maasweg 9, 2629 HZ Delft, The Netherlands

A DigitalMicrograph script *InsteaDMatic* has been developed for automated continuous rotation electron diffraction (cRED) data acquisition. *InsteaDMatic* coordinates functions of TEM goniometer and camera in order to tune up diffraction frame recording simultaneously with crystal rotation. Exploiting fast and automated data collection, the influence of the electron dose rate on the quality of cRED data was studied, and structural models obtained for two different parallel beam illumination modes (aperture selection and nanoprobe) have been compared.

**Abstract**     A DigitalMicrograph script *InsteaDMatic* has been developed to facilitate rapid automated continuous rotation electron diffraction (cRED) data acquisition. The script coordinates microscope functions, such as stage rotation, camera functions relevant for data collection, and stores the experiment metadata. The script is compatible with any microscope that can be controlled by DigitalMicrograph and has been tested on both JEOL and Thermo Fisher Scientific microscopes. A proof-of-concept has been performed through employing *InsteaDMatic* for data collection and structure determination of a ZSM-5 zeolite. The influence of illumination settings and electron dose rate on the quality of diffraction data, unit cell determination and structure solution has been investigated in order to optimize the data acquisition procedure.

## 1. Introduction

3D electron diffraction (3D ED) or microcrystal electron diffraction (MicroED) has been shown to be a powerful technique for structure determination of solids, which is especially advantageous for studies of micro- and nanocrystals. So far, hundreds of structures have been determined by 3D ED (Gemmi *et al.*, 2019), including zeolites (Jiang *et al.*, 2011; Martínez-Franco *et al.*, 2013; Guo *et al.*, 2015; Simancas *et al.*, 2016; Lee *et al.*, 2018; Bieseki *et al.*, 2018; Zhang *et al.*, 2018; Smeets *et al.*, 2019), metal-organic frameworks (Denysenko *et al.*, 2011; Feyand *et al.*, 2012; Wang, Rhauderwiek *et al.*, 2018; Lenzen *et al.*, 2019), pharmaceuticals (van Genderen *et al.*, 2018; Gruene *et al.*, 2018; Jones *et al.*, 2018; Brázda *et al.*, 2019), proteins (Nannenga *et al.*, 2014; de la Cruz *et al.*, 2017; Xu *et al.*, 2019; Xu *et al.*, 2019; Lanza *et al.*, 2019) and many others.

Data collection by 3D ED/MicroED was initially performed using a stepwise protocol, namely a set of electron diffraction patterns was recorded by tilting a crystal in a fixed angular step around an arbitrary crystallographic axis within the full range of the goniometer tilt (Kolb *et al.*, 2007; Zhang *et al.*, 2010; Shi *et al.*, 2013). Software packages dedicated for the stepwise 3D ED data collection and treatment were developed, known as Automated Diffraction Tomography, ADT (Kolb *et al.*, 2007) and Rotation Electron Diffraction, RED (Wan *et al.*, 2013). Soon after, data collection by continuous rotation of a crystal with a constant speed was proposed by several groups (Nederlof *et al.*, 2013; Nannenga *et al.*, 2014; Gemmi *et al.*, 2015; Yonekura *et al.*, 2015), leading to the development of a technique known as continuous rotation electron diffraction (cRED). cRED is performed by recording ED frames while continuously rotating the crystal along a goniometer axis with a constant speed. The basic hardware requirements for the TEM is only a single tilt sample holder and a camera. Hence, data collection can be performed on a wide variety of transmission electron microscopes. However, software control and synchronization of the TEM goniometer and the camera is required. Currently, only a limited number of software packages are designed to interface with both the camera and the microscope to collect multiple electron diffraction patterns simultaneously with crystal rotation. Many of them are commercial and/or closed source, e.g. *iTEM* from Olympus Soft Imaging Solutions (Gemmi *et al.*, 2015), *EPUd* (Thermo Fisher Scientific, 2019), *ParallEM* (Yonekura *et al.*, 2019), *eTasED* (Zhou *et al.*, 2019). Recently, a script for *SerialEM* (Mastronarde, 2005), a widely-used program in the cryo-electron microscopy community supporting electron microscopes and detectors from various manufacturers, has been used to enable large-scale MicroED data collection on Thermo Fisher Scientific (TFS) microscopes (de la Cruz *et al.*, 2019). Meanwhile, we have developed an open-source software platform *Instamatic* for electron crystallography needs, which is able to control both microscope and camera (Smeets *et al.*, 2018), and affords additional features such as crystal tracking through defocusing of diffraction pattern (Cichocka *et al.*, 2018; Wang *et al.*, 2019). Automation of the data collection through *Instamatic* allows reproducible results to be collected with minimal human efforts, especially for very large number of datasets. Currently, *Instamatic* is compatible with the Timepix detector (Amsterdam Scientific Instruments, The Netherlands) and the XF416/F416 cameras (Tietz Video and Image Processing Systems GmbH, Germany). However, additional developments are required for *Instamatic* to interface with other cameras. To the best of our knowledge, currently there is no flexible, cross-platform and easy-to-install software available for 3D ED data collection. Many existing software packages are optimized only for specific microscopes, which are installed in the working groups developing the software. Therefore, it is highly desirable to develop software that can interface and control a large variety of cameras and microscopes made by different manufacturers, and ensuring the hardware communications between them, even when they are controlled by separate computers. The software should be easy to set-up, straightforward to learn and user-friendly.

Here, we propose to employ DigitalMicrograph (DM, Digital Micrograph Gatan, Pleasanton, CA, U.S.A.) as a mediator controlling hardware interactions between microscope and camera. We developed a dedicated DM script, named *InsteaDMatic*, for automated cRED data collection. *InsteaDMatic* follows the same data collection workflow as described previously (Cichocka *et al.*, 2018) but communicates to both the microscope and camera *via* the DM interface. The benefit of this design philosophy is ease of installation and enhanced transferability, since the DM software is an integral part of a vast majority of electron microscopy systems nowadays. *InsteaDMatic* was first tested on our Themis Z (TFS) equipped with a Gatan OneView IS camera and JEM2100F (JEOL) with a Gatan Orius SC200D camera. Currently it has been successfully installed in more than 10 other labs, equipped by various types of TEMs (JEM2100F, JEM3100F, Titan, Talos) and different cameras (Ultrascan, Ceta, TVIPS). To demonstrate the capability of the script, we collected high-quality cRED data on a number of submicron-sized ZSM-5 zeolite crystals with up to 0.80 Å resolution allowing accurate structure determination. The resulting data statistics have been compared for crystals illuminated in selected area mode and in parallel nanoprobe mode. To highlight the advantages of the approach, parameters such as electron dose rate and monochromator focus have been tailored during the collection of cRED data.

**2. Experimental**

**2.1. Sample preparation**

Thoroughly ground ZSM-5 aluminosilicate zeolite powder was dispersed in ethanol followed by an ultrasonic bath treatment for 5 minutes. A drop of the suspension was applied to a lacey carbon grid (Cu150P from Okenshoji Co., Ltd) and dried in air for 10 minutes.

**2.2. Experimental setup**

The cRED experiments have been performed on a Themis Z microscope equipped with a Gatan OneView IS camera (4096 × 4096 pixels, pixel size: 15 µm) and a JEM 2100F equipped with a Gatan

Orius SC200D camera (2048 × 2048 pixels, pixel size: 7.4 µm). The OneView camera is well suited for cRED data acquisition, because it has essentially no readout dead time when in the movie mode. *In-situ* data capture mode with 1024 × 1024 pixels resolution (binning × 4) was employed. cRED data were collected using a single-tilt TFS holder (±40°), without applying a beam stopper. We found that Themis Z is very stable both electrically and mechanically, and the crystal tracking procedure described by Cichocka *et al.* (2018) is not a prerequisite for keeping the crystal centered in the electron beam during data collection. Before data acquisition, a standard TEM alignment routine was performed. All experiments were performed in the parallel illumination mode using a 50 µm C2 condenser aperture. The Z-height of the crystal was adjusted to the mechanical eucentric height in order to minimize its movement during tilting. Diffraction patterns were focused to obtain sharp spots in the diffraction mode. The rotation speed was 1.44°/s and the exposure time was 0.30 s/frame, leading to 0.432°/frame. A cRED dataset with a total rotation range of ~ 80° and 185 ED frames was collected in approximately 55 s.

Two different beam settings available on Themis Z were tested, namely selected area diffraction (SAED) and nanoprobe electron diffraction (NED) modes. In the SAED mode, a 40 µm SA aperture was inserted to limit the area used for diffraction, whereas in the NED mode the field of view was restricted by the beam size. Spot size 5 or 6 was usually used in the SAED mode, and spot size 11 in the NED mode. The electron dose on the specimen was controlled varying the monochromator focus.

For JEM 2100F equipped with a Gatan Orius SC200D detector (2048 × 2048 pixels, pixel size: 7.4 µm) the exposure time and rotation speed were set up to be 0.5 s/frame and 0.444 °/s, leading to 0.222°/frame and resulting in 209 frames within the total rotation range of 46.42° in 104.5 s. The relatively small tilt range was due to the limit of a single tilt holder for the microscope.

**2.3. Data processing and structure determination**

Diffraction images were collected in .TIFF format and converted to SMV format (.img) using the *process_DM* python script (Smeets, 2019). The collected frames were processed with the *XDS* software (Kabsch, 2010) for the spot-finding, unit cell determination, indexing, space-group assignment, data integration, scaling, and refinement. Previously determined lattice parameters and space group (Olson *et al.*, 1981) were used as an input, REFLECTING_RANGE_E.S.D. parameter in the XDS.INP file was set up to be 0.7 to include very sharp diffraction spots in the indexing procedure. Data statistics indicators provided in the output CORRECT.LP file were used further for data quality comparison. The reflection file for structure solution and refinement was obtained by merging several individual datasets from different crystals using the *XSCALE* sub-program. The structure was solved by *SHELXT* and refined by *SHELXL* (Sheldrick, 2008) using atomic structure factors for electrons (Doyle & Turner, 1968) with the help of *Olex2* software (Dolomanov *et al.*, 2009).

## 3. *InsteaDMatic* workflow

*InsteaDMatic* follows the data collection workflow described in (Cichocka *et al.*, 2018) using the continuous rotation method for electron diffraction (Arndt & Wonacott, 1977; Nederlof *et al.*, 2013; Nannenga *et al.*, 2014; Gemmi *et al.*, 2015). The same workflow has previously been implemented in Python in the program *Instamatic* (Smeets *et al.*, 2018). However, *Instamatic* requires additional development to interface different cameras.

On the camera computer, *InsteaDMatic* is run from DM and the GUI is shown in Figure 1. Settings for data collection (exposure, binning, *etc*.) are set through the camera panel in DM. When an experiment is started by pressing the "Start" button at the very bottom of the GUI, the script enters a waiting state where it constantly polls the current α tilt value. Once a change larger than a pre-defined threshold (angle activation threshold, typically 0.2°) is detected, data acquisition is initiated. The threshold also serves to eliminate any existing backlash in the α tilt direction. Rotation can then be initiated through any means available, either using the knobs, through the TEM user interface, or software. At present, the DM API does not allow fine control over the rotation speed of the goniometer, although this function is available on our microscope (Themis Z, TFS) through the *TEMScripting* interface, as well as other recent TFS/JEOL microscopes. To be able to control the rotation through DM, we implemented a custom Python script in *Instamatic* (Smeets, 2018) to synchronize rotation with data acquisition. The script establishes an interface with the TEM on the microscope computer and accepts connections over the network. A socket interface is then established using the program 'netcat'[1] on the camera computer through the DM function *LaunchExternalProcess*, which then communicates the requested rotation range and speed over the network to the microscope computer. Once rotation has been detected, data acquisition is initiated. The DM script hooks into the live view of the OneView camera, and then constantly copies front-most image to a pre-allocated "image buffer" whose size can be defined in the GUI of the script ("buffer size") and corresponds to the maximum number of frames that are expected to be collected. Whenever the live view is updated, DM fires an event called *DataValueChangedEvent*, which signals the script to copy the frame. The exposure time and binning are therefore defined through the DM interface, and not through the script. Data collection may be interrupted at any time by pressing the "Stop" button. There is also an automatic check for the completion of data collection, by monitoring the change of α tilt after every image cloning operation. When the change is equal to 0, the data collection loop breaks automatically. Finally, the script stores all relevant experimental metadata required for processing to a new directory, such as the rotation range, exposure time, camera length, etc. The image files are stored in the same directory in TIFF format, and can be converted to other desired formats (SMV and MRC) by running the *process_DM.py* script (Smeets, 2019).

---

[1] https://nmap.org/ncat/

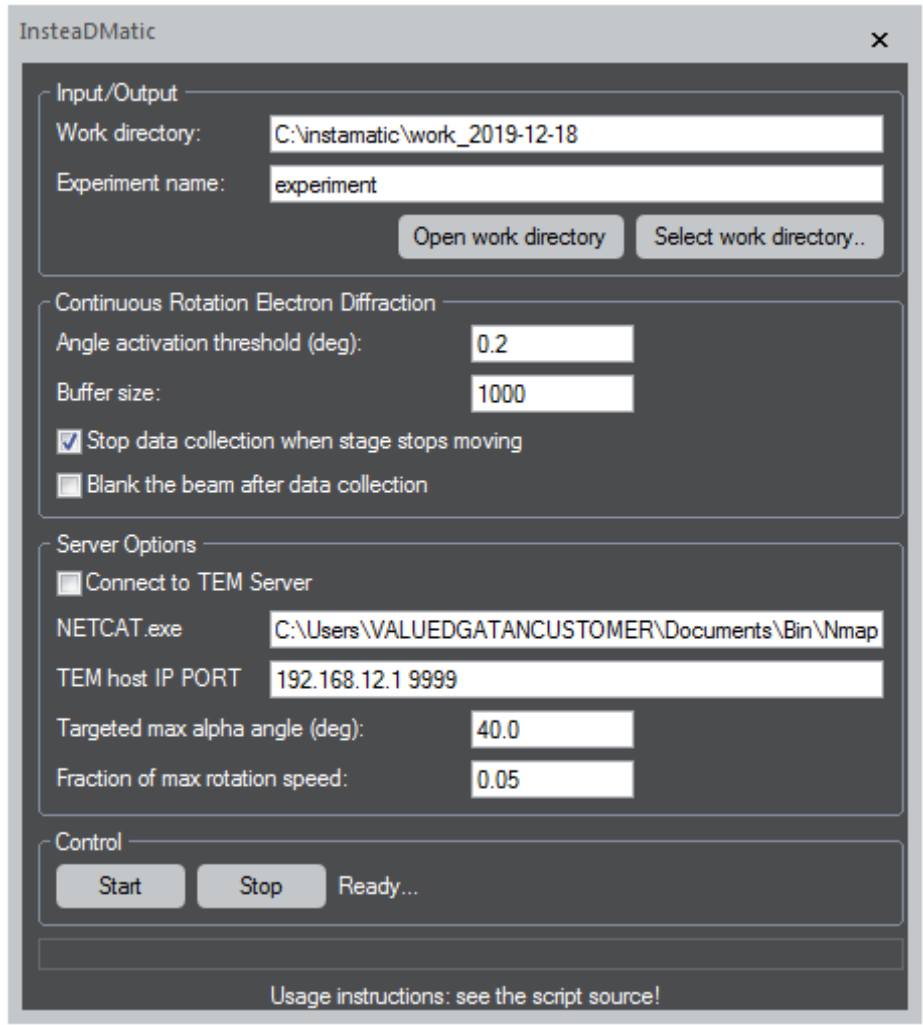

**Figure 1** The graphic interface of *InsteaDMatic*.

A flowchart of the workflow is shown in Figure 2. Detailed instruction of usage can be found from the script. The script is compatible with DM version 2.0 (which introduced the *DataValueChangedEvent*) or newer, and can be used with any Gatan camera that supports a streaming live view.

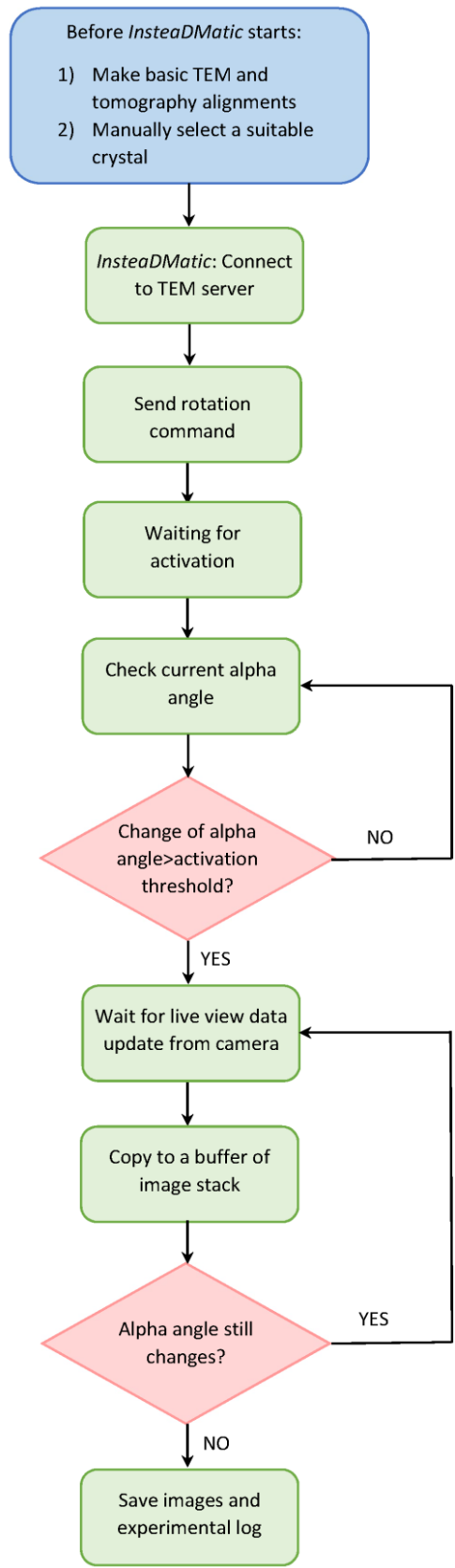

**Figure 2** Workflow for 3D ED data collection using *InsteaDMatic*. The blue box includes operations to be performed by a TEM operator, whereas the green and pink boxes show steps of the *InsteaDMatic* script protocol running automatically.

## 4. Application for structure determination of ZSM-5

A proof-of-concept has been performed through employing *InsteaDMatic* for data collection and further structure determination of a ZSM-5 aluminosilicate zeolite widely used in industry as a catalyst (Choi *et al.*, 2009; Ji *et al.*, 2017). ZSM-5 is relatively stable against electron beam damage, allowing multiple datasets to be collected from the same crystal. Consequently, a direct comparison of cRED data quality at different illumination settings becomes possible. ZSM-5 was previously used as a test sample for the assessment of data quality and accuracy of structure determination by rotation electron diffraction (RED) (Su *et al.*, 2014), cRED (Wang, Yang *et al.*, 2018), serial rotation electron diffraction (Wang *et al.*, 2019).

First, we tested *InsteaDMatic* on Themis Z with a Gatan One View CCD camera. A typical experiment was recorded in order to illustrate the procedure of cRED data acquisition, see Supporting Movie 1. The best Themis Z dataset demonstrated the completeness of more than 75% in the resolution shells ranging from 2.36 Å to 0.80 Å and $R_{meas}$ of 13.7% (see Table S1) enabling *ab initio* crystal structure solution from this one individual dataset. Unfortunately, the completeness of most individual datasets does not exceed 50% for the orthorhombic structure, and often only merged data can provide the correct structure (see below). We found that the OneView camera is well suited for experiments that require continuous read-out of the sensor. To check if the script would work on other cameras, we tested it on an Orius SC200D detector installed on a JEM 2100F. A "single-crystal" dataset collected over a rotation range of 46.42° reached the completeness of ~30% in the resolution shells from 2.36 Å to 0.80 Å, and exhibited the $R_{meas}$ of 26.1%. Due to the limited tilting capability of the microscope, the data completeness is low, prohibiting a correct crystal structure solution by direct methods, e.g., *Sir2014* (Burla *et al.*, 2015) or *SHELXT* (Sheldrick, 2008).

Traditionally, collection of electron diffraction data has been performed via diffraction area selection of a region of interest (ROI). However, the ROI selection can also be accomplished by adjusting the illumination settings. Almost parallel illumination with sub-micron beam diameter can be obtained either by Köhler illumination (Wu *et al.*, 2004; Meyer *et al.*, 2006; Benner *et al.*, 2011), or by inserting a small 10 µm C2 condenser aperture (Kolb *et al.*, 2007; Dwyer *et al.*, 2007). NED provides full control on the beam diameter and in principle allows collecting data on a smaller area with respect to SAED (Gemmi *et al.*, 2019). However, in the literature there is a lack of direct comparison of data quality collected on the same sample by cRED in SAED and NED modes. Here, an attempt has been made to reveal the difference between these modes using the same area of the sample for collecting diffraction data. In the SAED mode, a diffraction field of about 750 nm was selected by inserting the SA. In the NED mode the beam was condensed to illuminate the 750 nm area, and the electron dose rate was kept equal to those in the SAED mode by adjusting the monochromator focus. The two resulting datasets registered on the same isolated crystal are present in Table 1.

**Table 1** Typical parameters for data collection and data processing of individual datasets by XDS. Statistics in different resolution shells is given in Tables S2-S3.

|  | SAED dataset | NED dataset |
| --- | --- | --- |
| Spot size | 5 | 11 |
| Dose rate, e/Å$^2$s | 0.05 | 0.05 |
| Diffraction area, nm | 750 | 750 |
| Tilt range, ° | 39.64 to -40.00 | -39.71 to 40.00 |
| Tilt step, ° | 0.430 | 0.429 |
| Exposure time, s | 0.30 | 0.30 |
| Camera length, mm | 580 | 580 |
| Mono focus | 100.34 | 78.89 |
| Rotation speed, °/s | 1.441 | 1.434 |
| Total No. of reflections | 17224 | 17825 |
| No. of unique reflections | 2622 | 2692 |
| Completeness, % | 47.1 | 48.3 |
| Resolution cutoff, Å | 0.80 | 0.80 |
| $I/\sigma$ | 4.19 | 4.42 |
| $R_{obs}$, % | 20.8 | 21.7 |
| $R_{exp}$, % | 23.9 | 24.8 |
| $R_{meas}$, % | 22.8 | 23.9 |
| $CC_{1/2}$ | 98.7 | 98.1 |
| Unit cell parameters |  |  |
| a/Å | 20.38(4) | 20.56(4) |
| b/Å | 19.58(1) | 19.59(1) |
| c/Å | 13.21(2) | 13.18(2) |

Based on the previous crystallographic reports about the ZSM-5 crystal structure (Olson *et al.*, 1981; van Koningsveld *et al.*, 1987), the lattice parameters *a* = 20.022 Å, *b* = 19.899 Å, *c* = 13.383 Å and the space group *Pnma* (#62) were used as an input for XDS. Both SAED and NED datasets fit well with the expected orthorhombic structure and the refined unit cell parameters are close to the published values within the accuracy of the 3D ED method. Figure 3 shows the reconstructed reciprocal lattice of ZSM-5 based on the cRED data collected in the SAED mode, from Table 1.

research papers

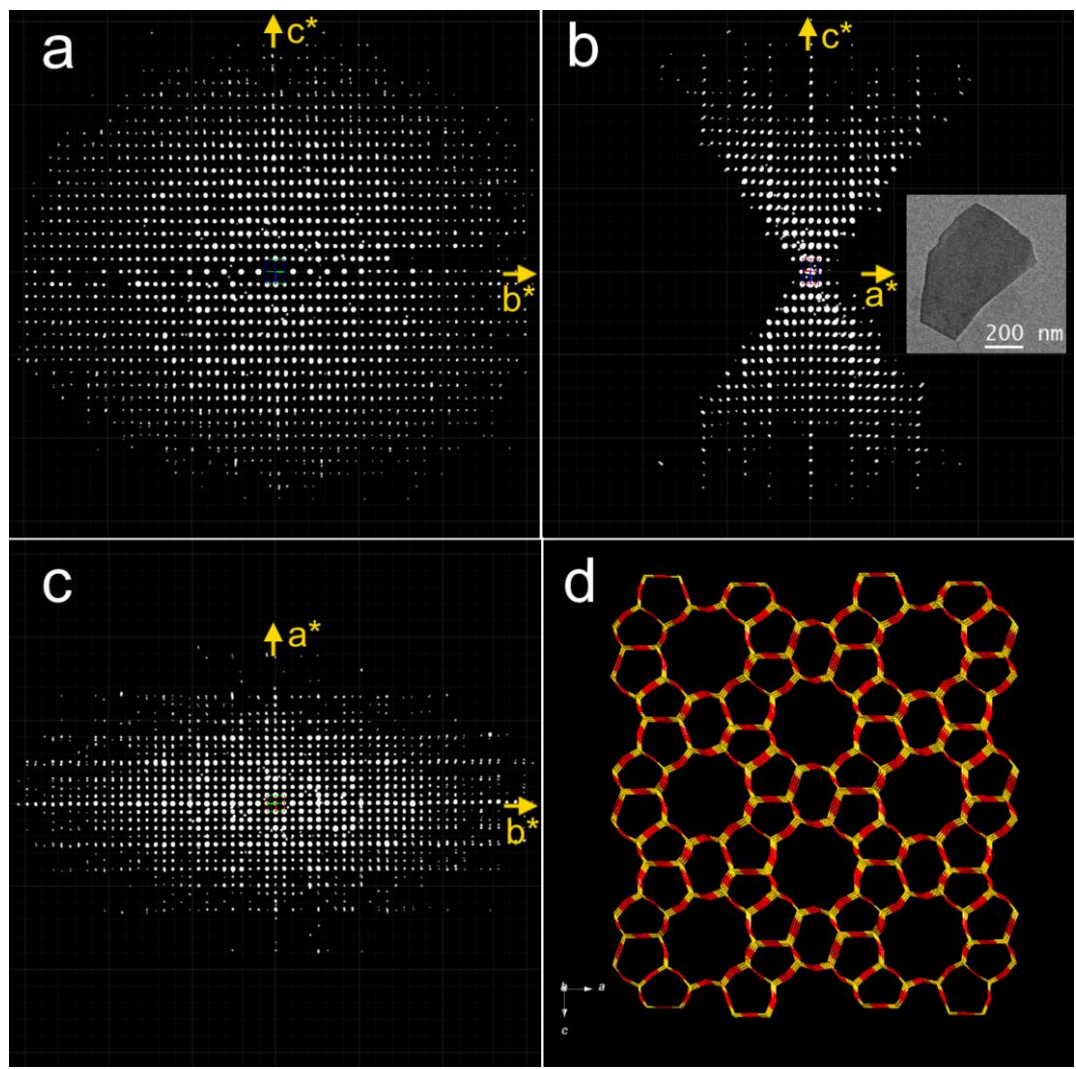

**Figure 3** (a-c) Typical 3D reciprocal lattice of ZSM-5 reconstructed and visualized by *REDp* (Wan *et al.*, 2013). The corresponding crystal image is shown as an inset in (b). (d) The wires/sticks ZSM-5 crystal structure representation.

Among factors affecting the cRED data quality, electron dose has an utmost importance. Our experiments have shown that the optimal electron dose rate range for ZSM-5 data acquisition is approximately between 0.03 e/Å$^2$s and 0.10 e/Å$^2$s (Figure 4). In the optimal range with no saturation, the higher the dose the better the *I/σ*. Excessive electron dose (>0.20 e/Å$^2$s) causes read-out biases of the OneView camera, whereas a low electron dose rate (<0.03 e/Å$^2$s) leads to significant deterioration of the signal-noise ratio and, as a consequence, to poor data statistics. Examples of the raw SAED/NED diffraction patterns collected at different electron doses are shown in Figure S1. Since XDS is relying upon the lowest measured intensities to guide the subtraction of the background, the scaling of the Bragg intensities as a function of resolution shells unavoidably leads to significant deteriorations of weak but still useful high-resolution signal, and consequently, to higher *R*-values in the high resolution shells (1.00 – 0.80 Å). For X-ray diffraction a common practice would be a truncating data at the resolution at which $R_{meas}$ remains below ~60% and $\langle I/\sigma \rangle$ is ~2 or higher. However, for electron diffraction, we found that including data out to a $CC_{1/2}$ value (Karplus & Diederichs, 2015) of ~70% leads to an improvement of the refined model even though the data at that resolution have a high $R_{meas}$.

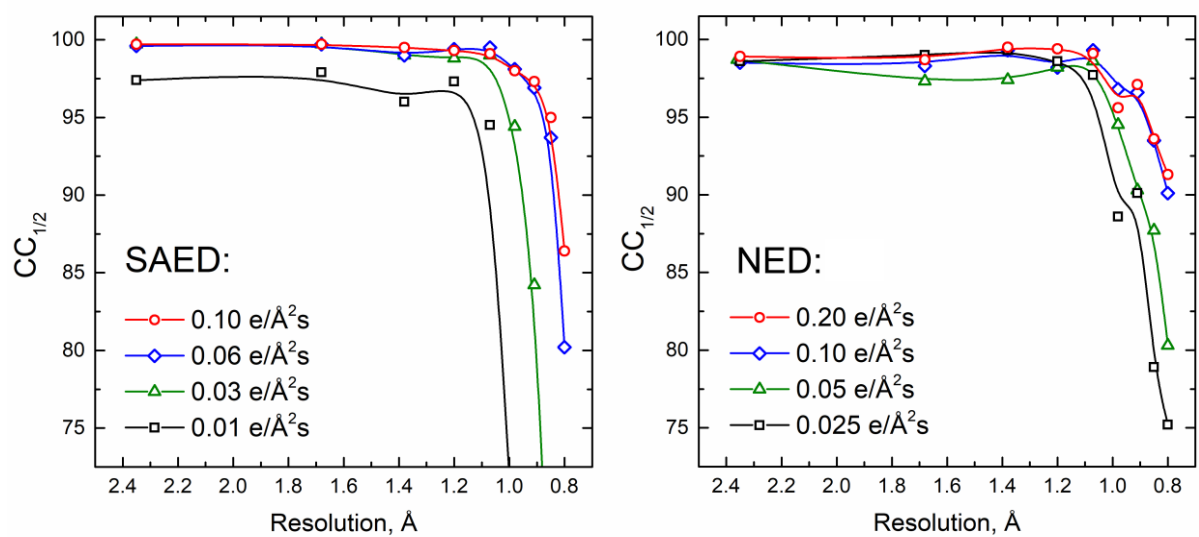

**Figure 4** Effect of the electron dose rate to $CC_{1/2}$ in different resolution shells. cRED data were collected with ROI selection either by selected-area aperture (SAED) or by nanoprobe illumination (NED). The ROI diameter was 750 nm. All SAED data were collected from the same ZSM-5 crystal sequentially, in the ascending order of the electron dose rate. All NED data were collected from a second crystal following the same procedure. The lines in the figure are a guide for the eye.

Another important factor for data collection is the stability of the *CompuStage*, since currently *InsteaDMatic* does not provide an opportunity to track the continuously rotating crystal during the data collection. We have shown that the specimen movement controlled by the *CompuStage* controller is smooth and crystal does not move out of the beam even without its position realignment during cRED data acquisition. In a typical experiment, we observed only a total drift of a few-nm for a 100 × 100 nm crystal rotated from -40 to 40°, accompanied by a jump of ~50 nm in the beginning of the rotation, see Supporting Movie 2. It should be noted that crystal drift becomes more severe at high tilt angles and ED intensities are also commonly systematically disrupted by uncompensated *Z*-height changes. It is therefore a better strategy to merge a number of datasets collected from different crystals in the range from -40 to 40°, instead of collecting cRED data from 1-2 crystals with a large tilt range, e.g. from -70 to 70°.

For the structure solution, five individual cRED datasets collected from different crystals in the SAED mode were merged, chosen by performing hierarchical cluster analysis based on 15 datasets using a home-developed software *edtools*[2]. For the NED data, 6 from 10 input cRED datasets were chosen for merging. Hierarchical cluster analysis helps to find structurally similar data with high correlation coefficients between scaled diffraction intensities and to reach high completeness by merging only few datasets (Wang *et al.*, 2019). However, it is worth to note that simple averaging of unit-cell parameters obtained from individual 3D ED datasets may result in irrelevant interatomic distances in the final structure; hence, the lattice parameters of standard ZSM-5 (as-made ZSM-5, determined by single-crystal XRD (van Koningsveld *et al.*, 1987)) were used for the structure solution and refinement, see the International Zeolite Association (IZA) Database.

The structure of ZSM-5 can be solved using either *Sir2014* (Burla *et al.*, 2015) direct space or *SHELXT* dual space methods (Sheldrick, 2008). We note that a minimal $I/\sigma$ signal-to-noise ratio of ca. 2 (at 1.0 Å resolution limit) is required for revealing the framework of ZSM-5 by means of direct methods, whereas dual space methods are not so sensitive to the $I/\sigma$ ratio. There are 38 symmetry-independent atoms in the ZSM-5 structure, of which 12 are Si atoms and 26 are O atoms. There are four O atoms located at special positions. The atomic positions of all 12 Si- and 24 O-atoms can been successfully found using both NED and SAED data, and used as an initial structural model. The details of structure refinement are provided in Table 2. Anisotropic refinement of the NED model leads to $R_1$ = 0.1758, GooF = 1.61, Si–O bond lengths in the range of 1.47 – 1.63 Å, and O–Si–O angles in the range of 104.9 – 117.4°, with no additional restraints applied. For the SAED data, the refinement converged with $R_1$ = 0.1992, GooF = 1.58, and two restraints were applied to keep Si–O bond lengths reasonable, arguing for somewhat better data quality of the merged NED dataset. In full agreement with the SCXRD model (Olson *et al.*, 1981; van Koningsveld *et al.*, 1987), the framework structure of ZSM-5 obtained from cRED data has a three-dimensional channel system with 10-ring straight channels of 5.4 × 5.6 Å in diameter running parallel to [010] and 10-ring sinusoidal channels of 5.1 × 5.4 Å in diameter running parallel to [100].

A comparison with the reference model obtained from SCXRD (van Koningsveld *et al.*, 1987) was carried out using COMPSTRU program (Flor *et al.*, 2016). All deviations of atomic positions between the reference ZSM-5 structure and those determined from cRED data are listed in Table S6. It has been shown that the deviations for the model obtained from the merged NED dataset (on average 0.03(1) Å for Si and 0.05(2) Å for O) are lower than those for the model obtained from the merged SAED dataset (0.05(1) Å for Si and 0.07(3) Å for O). The accuracy of the models is comparable to that obtained from our previous studies using single cRED datasets collected in SAED mode on a JEM-2100 LaB$_6$ microscope (Wang, Yang *et al.*, 2018). However, in the present study, no restraints were needed for the structure refinements, while in the previous cases it was necessary to apply restraints on all Si—O distances in order to keep a reasonable geometry of the structure model (Wang, Yang *et al.*, 2018).

---

[2] https://github.com/stefsmeets/edtools

**Table 2** Selected crystallographic data for merged ZSM-5 datasets. Space group *Pnma* (#62), unit cell parameters $a$ = 20.022(4) Å, $b$ = 19.899(4) Å, $c$ = 13.383(3) Å, electron wavelength $\lambda$ = 0.019 Å. Statistics in different resolution shells is given in Tables S4-S5

|  | SAED | NED |
| --- | --- | --- |
| Datasets merged | 5 | 6 |
| Total No. of reflections | 61596 | 65672 |
| No. of unique reflections | 5159 | 5299 |
| No. of reflections with I > 2σ(I) | 2854 | 3903 |
| $R_{int}$ | 0.3082 | 0.2282 |
| Completeness, % | 95.8 | 98.2 |
| Resolution cutoff, Å | 0.80 | 0.80 |
| No. parameters | 332 | 332 |
| No. restraints | 2 | 0 |
| $R_1$ (I > 2σ(I)) | 0.1992 | 0.1758 |
| $R_1$ (all data) | 0.2612 | 0.1997 |
| GooF | 1.58 | 1.61 |

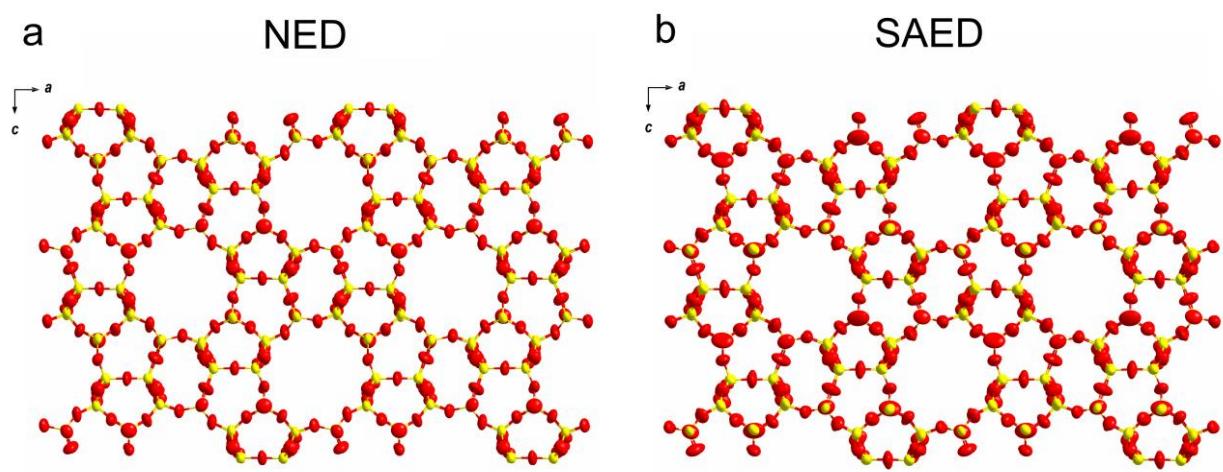

**Figure 5** Framework structure of ZSM-5 viewed along *b*-axis refined using (a) NED and (b) SAED data showing anisotropic atomic displacement parameters for Si (yellow) and O (red) atoms. The ellipsoids are drawn at the 50% probability level.

In contrast to the SAED mode, where the ROI to be used for data collection is pre-defined by the selected area aperture size, the NED mode provides higher flexibility in adjusting the size of the area to be illuminated, hence, to fitting the size of each individual crystal so that the background in the ED frames is largely eliminated. This may be highly beneficial for studies of beam sensitive materials since it paves an avenue for tailoring of the electron dose received by a specimen in a controllable manner. NED parallel illumination also provides slightly more accurate structure models as compared to SAED mode.

## 5. Conclusion

A new custom DigitalMicrograph script named *InsteaDMatic* has been developed to facilitate rapid automated continuous rotation electron diffraction data acquisition. *InsteaDMatic* has been successfully installed and operated on JEOL and Thermo Fisher Scientific microscopes utilizing DigitalMicrograph for control over the instrument and camera. The script was employed for data collection and structural determination of the ZSM-5 zeolite framework. Dose rate between 0.03 e/Å$^2$s and 0.10 e/Å$^2$s was found to be optimal for obtaining high quality data with up to 0.80 Å resolution. Positions of the Si and O atoms in ZSM-5 can be found within an accuracy better than 0.03 Å and 0.05 Å, respectively, compared to those obtained by single-crystal XRD data. Both SAED and NED beam settings deliver an accurate structural model, provided that the beam and the stage are stable during the goniometer rotation. Varying the monochromator focus offers an additional degree of freedom for tailoring the electron dose, which is especially relevant in the NED mode. We anticipate that the present research will contribute to the development of widely applicable routine for the structure determination of micro- and nanocrystals by 3D ED.

The *InsteaDMatic* script described in this article is available from
https://github.com/stefsmeets/InsteaDMatic

During the manuscript preparation, the *InsteaDMatic* script was successfully installed and tested in more than 10 electron microscopy labs worldwide, and we highly acknowledge the feedback that we are receiving from them.

**Acknowledgements**

The project is supported by the Swedish Research Council (VR, 2017-04321 to XZ, 2017-05333 to HX), the Knut and Alice Wallenberg Foundation through the project grant 3DEM-NATUR (2012.0112) and the Swiss National Science Foundation (SNSF, 177761 to SS).

# Supporting information

**Table S1** Comparison of cRED data from Themis Z with One View camera and JEM2100F with Orius SC200D camera.

|  | Themis Z / One View | JEOL2100F / Orius SC200D |
|---|---|---|
| Spot size | 5 | 1 |
| Dose rate, e/Å$^2$s | 0.05 | 0.084 |
| Diffraction area, nm | 750 | 1200 |
| Tilt range, ° | 29.77 to -29.99 | -22.48 to 23.94 |
| Oscillation angle, ° | 0.424 | 0.222 |
| Exposure time, s | 0.30 | 0.50 |
| Acquisition time per frame, s | 0.30 | 0.50 |
| Camera length, mm | 580 | 800 |
| Rotation speed, °/s | 1.421 | 0.444 |
| Rotation axis, ° | -171 | -42.5 |
| Total No. of reflections | 12055 | 11042 |
| No. of unique reflections | 4276 | 2086 |
| Completeness, % | 77.7 | 34.5 |
| Resolution cutoff, Å | 0.80 | 0.80 |
| I/$\sigma$ | 4.19 | 3.08 |
| $R_{obs}$, % | 11.0 | 23.3 |
| $R_{exp}$, % | 12.4 | 30.0 |
| $R_{meas}$, % | 13.7 | 26.1 |
| $CC_{1/2}$ | 99.1 | 98.6 |
| Unit cell parameters |  |  |
| a/Å | 19.93(2) | 20.68(2) |
| b/Å | 19.48(3) | 20.49(2) |
| c/Å | 13.461(9) | 13.75(1) |

**Table S2** Statistics of an individual cRED dataset collected in SAED mode on Themis Z in different resolution shells.

| Resolution limit | #obs | #uniq | #pos | comp % | $R_{obs}$ % | $R_{exp}$ % | #comp | $I/\sigma$ | $R_{meas}$ % | $CC_{1/2}$ |
|---|---|---|---|---|---|---|---|---|---|---|
| 2.35 | 605 | 107 | 233 | 45.9 | 12.2 | 13.2 | 601 | 10.91 | 13.5 | 98.3 |
| 1.68 | 1101 | 181 | 382 | 47.4 | 16.0 | 14.9 | 1097 | 8.24 | 17.6 | 98.0 |
| 1.38 | 1460 | 233 | 496 | 47.0 | 17.0 | 16.0 | 1452 | 7.41 | 18.7 | 98.5 |
| 1.20 | 1789 | 276 | 574 | 48.1 | 19.2 | 19.1 | 1785 | 6.27 | 21.2 | 97.9 |
| 1.07 | 2103 | 312 | 650 | 48.0 | 21.8 | 22.0 | 2098 | 5.47 | 23.7 | 97.9 |
| 0.98 | 2315 | 344 | 719 | 47.8 | 43.8 | 52.1 | 2300 | 3.09 | 47.6 | 93.9 |
| 0.91 | 2502 | 374 | 786 | 47.6 | 48.0 | 72.4 | 2489 | 2.51 | 52.2 | 92.5 |
| 0.85 | 2724 | 400 | 832 | 48.1 | 70.5 | 117.0 | 2714 | 1.71 | 76.5 | 90.1 |
| 0.80 | 2625 | 395 | 890 | 44.4 | 89.3 | 173.9 | 2614 | 1.21 | 97.0 | 88.7 |
| total | 17224 | 2622 | 5562 | 47.1 | 20.8 | 23.9 | 17150 | 4.19 | 22.8 | 98.5 |

**Table S3** Statistics of an individual cRED dataset collected in NED mode on Themis Z in different resolution shells.

| Resolution limit | #obs | #uniq | #pos | comp % | $R_{obs}$ % | $R_{exp}$ % | #comp | $I/\sigma$ | $R_{meas}$ % | $CC_{1/2}$ |
|---|---|---|---|---|---|---|---|---|---|---|
| 2.36 | 653 | 112 | 234 | 47.9 | 13.2 | 15.6 | 651 | 9.74 | 14.6 | 98.7 |
| 1.68 | 1149 | 182 | 384 | 47.4 | 17.5 | 16.8 | 1146 | 8.26 | 19.2 | 97.3 |
| 1.38 | 1500 | 239 | 490 | 48.8 | 21.4 | 18.3 | 1495 | 7.15 | 23.5 | 95.4 |
| 1.20 | 1845 | 275 | 578 | 47.6 | 22.4 | 21.0 | 1843 | 6.61 | 24.8 | 98.2 |
| 1.07 | 2092 | 319 | 649 | 49.2 | 23.0 | 22.9 | 2079 | 5.71 | 25.1 | 98.6 |
| 0.98 | 2382 | 351 | 721 | 48.7 | 40.9 | 55.0 | 2376 | 3.48 | 44.8 | 94.5 |
| 0.91 | 2600 | 380 | 789 | 48.2 | 44.8 | 69.2 | 2595 | 3.10 | 49.1 | 90.3 |
| 0.85 | 2782 | 409 | 829 | 49.3 | 57.6 | 101.3 | 2767 | 2.25 | 62.9 | 90.7 |
| 0.80 | 2822 | 425 | 899 | 47.3 | 65.2 | 144.0 | 2810 | 1.49 | 71.1 | 93.7 |
| total | 17825 | 2692 | 5573 | 48.3 | 21.7 | 24.8 | 17762 | 4.42 | 23.9 | 98.1 |

**Table S4** Statistics of SAED dataset merged from 5 crystals in different resolution shells.

| Resolution limit | #obs | #uniq | #pos | comp % | $R_{obs}$ % | $R_{exp}$ % | #comp | $I/\sigma$ | $R_{meas}$ % | $CC_{1/2}$ |
|---|---|---|---|---|---|---|---|---|---|---|
| 3.58 | 571 | 61 | 67 | 91.0 | 21.8 | 27.3 | 571 | 7.64 | 23.4 | 94.2 |
| 2.53 | 1051 | 104 | 110 | 94.5 | 25.4 | 27.1 | 1051 | 7.35 | 26.7 | 95.5 |
| 2.07 | 1479 | 143 | 150 | 95.3 | 29.5 | 27.7 | 1479 | 6.91 | 31.2 | 97.4 |
| 1.79 | 1796 | 164 | 175 | 93.7 | 23.7 | 27.7 | 1796 | 6.77 | 25.1 | 98.2 |
| 1.60 | 1927 | 179 | 189 | 94.7 | 32.6 | 28.6 | 1927 | 6.06 | 34.6 | 88.9 |
| 1.46 | 2228 | 201 | 208 | 96.6 | 32.7 | 30.5 | 2226 | 5.71 | 34.3 | 98.4 |
| 1.35 | 2641 | 233 | 241 | 96.7 | 34.1 | 29.9 | 2640 | 5.22 | 36.0 | 94.3 |
| 1.26 | 2755 | 238 | 244 | 97.5 | 50.8 | 33.7 | 2752 | 4.99 | 53.5 | 81.3 |
| 1.19 | 2995 | 256 | 267 | 95.9 | 34.2 | 33.0 | 2994 | 5.15 | 36.1 | 93.8 |
| 1.13 | 3165 | 269 | 277 | 97.1 | 40.7 | 35.2 | 3163 | 4.85 | 42.7 | 93.4 |
| 1.08 | 3373 | 278 | 289 | 96.2 | 48.2 | 43.5 | 3370 | 4.07 | 50.8 | 90.9 |
| 0.99 | 3883 | 304 | 318 | 95.6 | 93.8 | 101.8 | 3881 | 2.82 | 98.0 | 89.1 |
| 0.92 | 4093 | 327 | 334 | 97.9 | 113.9 | 131.4 | 4089 | 2.45 | 119.2 | 86.6 |
| 0.89 | 4500 | 350 | 366 | 95.6 | 134.7 | 160.6 | 4498 | 2.30 | 140.6 | 86.2 |
| 0.84 | 4823 | 371 | 380 | 97.6 | 251.8 | 337.5 | 4819 | 1.74 | 262.5 | 61.4 |
| 0.82 | 4732 | 366 | 381 | 96.1 | 296.4 | 405.5 | 4729 | 1.53 | 309.0 | 55.8 |
| 0.80 | 3357 | 337 | 390 | 86.4 | 255.0 | 394.5 | 3347 | 1.35 | 267.6 | 69.2 |
| total | 61596 | 5159 | 5386 | 95.8 | 33.0 | 33.8 | 61551 | 3.56 | 34.9 | 94.8 |

**Table S5** Statistics of NED dataset merged from 6 crystals in different resolution shells.

| Resolution limit | #obs | #uniq | #pos | comp % | $R_{obs}$ % | $R_{exp}$ % | #comp | $I/\sigma$ | $R_{meas}$ % | $CC_{1/2}$ |
|---|---|---|---|---|---|---|---|---|---|---|
| 3.58 | 705 | 67 | 68 | 98.5 | 20.0 | 23.6 | 705 | 9.53 | 21.2 | 98.0 |
| 2.53 | 1302 | 112 | 112 | 100.0 | 17.0 | 24.0 | 1302 | 9.00 | 17.9 | 98.5 |
| 2.07 | 1775 | 146 | 146 | 100.0 | 19.6 | 24.5 | 1775 | 8.59 | 20.6 | 97.7 |
| 1.79 | 2188 | 175 | 175 | 100.0 | 22.3 | 24.8 | 2188 | 7.81 | 23.6 | 97.5 |
| 1.60 | 2507 | 196 | 196 | 100.0 | 21.4 | 25.9 | 2507 | 7.44 | 22.5 | 98.5 |
| 1.46 | 2690 | 203 | 205 | 99.0 | 29.8 | 27.4 | 2690 | 6.98 | 31.2 | 99.6 |
| 1.35 | 3066 | 232 | 233 | 99.6 | 23.4 | 26.5 | 3066 | 6.84 | 24.6 | 84.3 |
| 1.27 | 3422 | 255 | 255 | 100.0 | 28.2 | 29.5 | 3422 | 6.13 | 29.5 | 98.9 |
| 1.19 | 3528 | 258 | 258 | 100.0 | 26.2 | 29.6 | 3528 | 6.38 | 27.4 | 98.8 |
| 1.13 | 3881 | 287 | 288 | 99.7 | 27.7 | 30.7 | 3881 | 5.83 | 28.9 | 96.6 |
| 1.08 | 4016 | 283 | 285 | 99.3 | 38.8 | 39.1 | 4016 | 5.22 | 40.4 | 98.8 |
| 0.99 | 3804 | 324 | 325 | 99.7 | 53.3 | 51.6 | 3804 | 3.73 | 55.9 | 95.3 |
| 0.92 | 4069 | 341 | 341 | 100.0 | 54.4 | 56.9 | 4069 | 3.41 | 57.1 | 92.7 |
| 0.89 | 4268 | 358 | 358 | 100.0 | 67.0 | 65.6 | 4268 | 2.89 | 70.1 | 92.3 |
| 0.84 | 4731 | 395 | 395 | 100.0 | 90.4 | 103.9 | 4731 | 2.27 | 94.7 | 84.3 |
| 0.82 | 4490 | 383 | 388 | 98.7 | 91.6 | 114.8 | 4490 | 2.18 | 95.9 | 95.7 |
| 0.80 | 3234 | 310 | 394 | 78.7 | 91.8 | 112.8 | 3227 | 1.83 | 96.6 | 83.7 |
| total | 65672 | 5299 | 5397 | 98.2 | 24.0 | 27.7 | 65665 | 4.56 | 25.2 | 97.4 |

**Table S6** Deviations of atomic positions between the reference ZSM-5 structure (van Koningsveld *et al.*, 1987) and those determined from cRED data collected in SAED/NED modes. Fractional atomic coordinates for the reference ZSM-5 structure determined by SCXRD are given in Table S7 (as-made ZSM-5, space group *Pnma*, $a$ = 20.022(4) Å, $b$ = 19.899(4) Å, $c$ = 13.383(3) Å, see the International Zeolite Association (IZA) Database).

| Atom | Atomic displacement SAED, Å | Atomic displacement NED, Å |
| --- | --- | --- |
| Si1 | 0.064 | 0.027 |
| Si2 | 0.063 | 0.042 |
| Si3 | 0.043 | 0.032 |
| Si4 | 0.021 | 0.027 |
| Si5 | 0.021 | 0.021 |
| Si6 | 0.063 | 0.007 |
| Si7 | 0.046 | 0.014 |
| Si8 | 0.051 | 0.022 |
| Si9 | 0.053 | 0.030 |
| Si10 | 0.051 | 0.036 |
| Si11 | 0.055 | 0.035 |
| Si12 | 0.044 | 0.016 |
| O1 | 0.074 | 0.050 |
| O2 | 0.105 | 0.053 |
| O3 | 0.022 | 0.037 |
| O4 | 0.067 | 0.051 |
| O5 | 0.032 | 0.028 |
| O6 | 0.066 | 0.025 |
| O7 | 0.070 | 0.001 |
| O8 | 0.053 | 0.029 |
| O9 | 0.041 | 0.026 |
| O10 | 0.101 | 0.082 |
| O11 | 0.011 | 0.047 |
| O12 | 0.122 | 0.124 |
| O13 | 0.098 | 0.073 |
| O14 | 0.097 | 0.032 |
| O15 | 0.075 | 0.060 |
| O16 | 0.046 | 0.028 |
| O17 | 0.081 | 0.048 |
| O18 | 0.095 | 0.049 |
| O19 | 0.070 | 0.043 |
| O20 | 0.045 | 0.030 |
| O21 | 0.072 | 0.036 |
| O22 | 0.023 | 0.073 |
| O23 | 0.070 | 0.080 |
| O24 | 0.068 | 0.060 |
| O25 | 0.075 | 0.065 |
| O26 | 0.092 | 0.029 |
| **<Si> average** | **0.05(2)** | **0.03(1)** |
| **<O> average** | **0.07(3)** | **0.05(2)** |

**Table S7** Fractional atomic coordinates for the reference ZSM-5 structure (van Koningsveld *et al.*, 1987).

| Atom | $x$ | $y$ | $z$ |
|---|---|---|---|
| Si1 | 0.4224 | 0.0565 | −0.3360 |
| Si2 | 0.3072 | 0.0277 | −0.1893 |
| Si3 | 0.2791 | 0.0613 | 0.0312 |
| Si4 | 0.1221 | 0.0630 | 0.0267 |
| Si5 | 0.0713 | 0.0272 | −0.1855 |
| Si6 | 0.1864 | 0.0590 | −0.3282 |
| Si7 | 0.4227 | −0.1725 | −0.3272 |
| Si8 | 0.3078 | −0.1302 | −0.1855 |
| Si9 | 0.2755 | −0.1728 | 0.0311 |
| Si10 | 0.1206 | −0.1731 | 0.0298 |
| Si11 | 0.0704 | −0.1304 | −0.1820 |
| Si12 | 0.1871 | −0.1733 | −0.3193 |
| O1 | 0.3726 | 0.0534 | −0.2442 |
| O2 | 0.3084 | 0.0587 | −0.0789 |
| O3 | 0.2007 | 0.0592 | 0.0289 |
| O4 | 0.0969 | 0.0611 | −0.0856 |
| O5 | 0.1149 | 0.0541 | −0.2763 |
| O6 | 0.2435 | 0.0553 | −0.2460 |
| O7 | 0.3742 | −0.1561 | −0.2372 |
| O8 | 0.3085 | −0.1552 | −0.0728 |
| O9 | 0.1980 | −0.1554 | 0.0288 |
| O10 | 0.0910 | −0.1614 | −0.0777 |
| O11 | 0.1169 | −0.1578 | −0.2694 |
| O12 | 0.2448 | −0.1594 | −0.2422 |
| O13 | 0.3047 | −0.0510 | −0.1866 |
| O14 | 0.0768 | −0.0519 | −0.1769 |
| O15 | 0.4161 | 0.1276 | −0.3896 |
| O16 | 0.4086 | −0.0017 | −0.4136 |
| O17 | 0.4020 | −0.1314 | −0.4239 |
| O18 | 0.1886 | 0.1298 | −0.3836 |
| O19 | 0.1940 | 0.0007 | −0.4082 |
| O20 | 0.1951 | −0.1291 | −0.4190 |
| O21 | −0.0037 | 0.0502 | −0.2080 |
| O22 | −0.0040 | −0.1528 | −0.2078 |
| O23 | 0.4192 | −0.2500 | −0.3540 |
| O24 | 0.1884 | −0.2500 | −0.3538 |
| O25 | 0.2883 | −0.2500 | 0.0579 |
| O26 | 0.1085 | −0.2500 | 0.0611 |

**Figure S1** Typical diffraction patterns of ZSM-5 collected on Themis Z / One View at different electron dose rates. Dotted rings indicate 0.8 Å resolution.

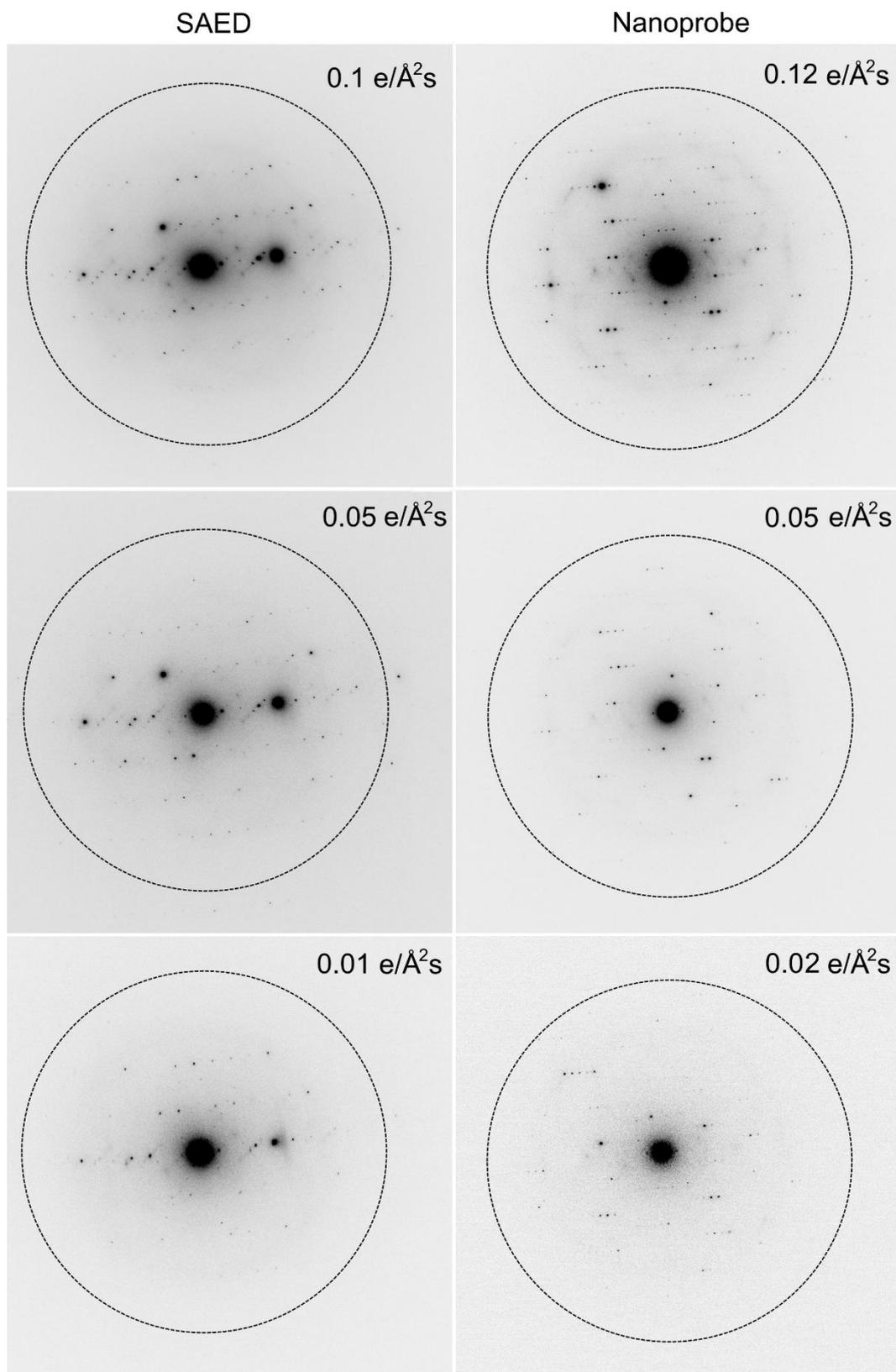

**Supporting movie S1**

https://stockholmuniversity.box.com/s/96cfu9kspo4vohuzynz092o47cpj0722

**Supporting movie S2**

https://stockholmuniversity.box.com/s/f161l0mb015o97d5qurdsharahpcpfzz